\documentclass[aps,twocolumn,prb,showpacs,groupedaddress]{revtex4}
\usepackage{graphicx}

\begin{document}
\bibliographystyle{apsrev}

\preprint{cond-mat/0504090}

\title[Photoacoustic wave in Pb]%
{Photoacoustic wave propagating from normal into superconductive phases 
in Pb single crystals}

\author{Masanobu Iwanaga}
\email[Electronic address: ]{iwanaga@phys.tohoku.ac.jp}
\affiliation{Department of Physics, Graduate School 
of Science, Tohoku University, Sendai 980-8578, Japan}

\date{\today}

\begin{abstract}
Photoacoustic (PA) wave has been examined in a superconductor 
of the first kind, Pb single crystal. The PA wave is induced by 
optical excitation of electronic state and propagates from normal 
into superconductive phases below T$_{\rm C}$. 
It is clearly shown by wavelet analysis that the measured PA wave 
includes two different components. The high-frequency component 
is MHz-ultrasonic and the relative low-frequency one is induced by 
thermal wave. The latter is observed in a similar manner irrespective 
of T$_{\rm C}$. On the other hand, the MHz-frequency component 
is obviously enhanced below T$_{\rm C}$. 
The behavior is reproduced by the change of attenuation of 
longitudinal ultrasonic wave and is consistent with BCS theory. 
\end{abstract}

\maketitle

\section*{INTRODUCTION}
Photoacoustic (PA) spectroscopy has the special advantage of 
analyzing thermal and elastic signals induced by photoexcitation, 
and has been widely applied to gas, liquid, and solids.~\cite{Patel,Tam} 
Since the PA signal detected by piezoelectric transducer (PZT) is 
thermoelastic,~\cite{Patel} PA spectroscopy is considered 
effective to examine phase transitions. 
Indeed, PA signals around 
phase-transition point were theoretically studied,~\cite{Korpiun,Etxe} 
and several observations for first- and second-order transitions have been 
reported so far.~\cite{Etxe,Somasun,Kojima,Iwanaga_e} 
In the previous reports, the change of PA signal has not been 
definitely attributed to the change of a physical quantity. 
This is because several physical parameters can contribute to the 
change of PA signal. 
In addition, since the PA signal was usually picked up by a lockin detector, 
the information is expressed by only two values, the amplitude and phase. 
Thus, the PA technique has had the advantage and disadvantage: 
the access to several 
physical quantities and the ambiguity in the interpretation. 
To extract more information in PA measurement, it is probably significant 
to examine PA wave itself. 
Generally, PA waves are generated from the heat source 
which results from nonradiative energy relaxation of photoexcited electrons. 
From the generation process, PA waves are regarded as the 
thermal and/or elastic wave associated with energy dissipation. 
The PA waves are expected to include the modes peculiar to the medium. 
If it is true, the analysis in the time and frequency domains will be 
helpful to make the physical properties clear; 
however, such a kind of study has hardly been reported. 

In the PA studies to date, superconductive transitions have not been 
examined to our knowledge. As known prevailingly, 
the transitions are second-order without any crystallographic 
transition and are responsible for drastic change of electric 
conductivity. The transition has been most extensively investigated in 
various physical properties such as magnetic, thermal, and ultrasonic 
properties. Thus, the superconductor described 
by BCS theory~\cite{Bardeen} seems suitable to test physical quantities 
detected in PA measurement. In this study, 
it is an aim to clarify the properties of PA wave and signals. 
Moreover, it is expected to reveal how one can analyze superconductive 
transition with PA spectroscopy. When photoexcitation induces electronic 
interband transition, it destroys 
superconductive phase due to the energy far larger than the energy gap 
in the phase. The effect is also discussed. 

Concretely, Pb single crystal is explored in this study. 
The crystal is a superconductor of the first kind described by 
the strong electron-phonon coupled BCS theory; 
the superconductive phase has been closely 
investigated with far infrared spectroscopy,~\cite{Richards} 
electron tunneling technique,~\cite{Giaever} ultrasonic pulse-echo 
technique,~\cite{Bommel,Deaton,Tittmann,Fate} and so on. 
The crystal has the 
critical temperature T$_{\rm C}$ of 7.22 K (Ref.\ \onlinecite{Kittel}). 
Thus, various material 
parameters have been obtained so far. 

\section*{EXPERIMENT}

\begin{figure}
\includegraphics[width=8.0cm,clip]{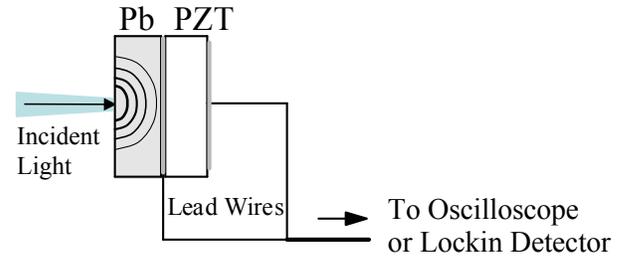}%
\caption{Experimental configuration. Photoacoustic (PA) wave is drawn 
schematically in the Pb single crystal. PZT denotes piezoelectric 
transducer. Experimental parameters and conditions are 
described in the text.}
\label{fig1}
\end{figure}

The Pb single crystal has the purity more than 99.999\% and is the size of 
5$\times$5$\times$1 mm$^3$; the plane of 5$\times$5 mm$^2$ is 
(100) plane and the thickness is 1 mm. The 
PZT of the same size as the Pb single crystal was used in the PA 
measurement; the PZT has the resonant frequency of 4.00 MHz and the Curie 
point at 603 K. As drawn in 
Fig.\ \ref{fig1}, the (100) plane of Pb crystal was firmly attached 
to the 5$\times$5 mm$^2$ plane of the PZT with conductive organic paste. 
Since the lead wires were also attached as shown in Fig.\ \ref{fig1}, 
the detected voltage is proportional to the stress along the 
thickness direction, and the detected PA wave is bulk wave which propagates 
through the crystal. In the present configuration, the bulk wave generally 
includes the wave along off-thickness direction. 
The specimen and PZT were set in a He-flow 
cryostat equipped with a temperature controller. 

Incident 2.33-eV light in measuring PA wave was second harmonics 
of a YAG (yttrium-alminium-garnet) laser and was injected onto (100) 
plane; the pulse width was 5 ns, and the repetition was 10 Hz. 
The incident light is strongly absorbed by Pb single crystals because of 
the electronic interband transition; the absorption length is 26 
nm (Ref.\ \onlinecite{Liljenvall}). Therefore, the incident photons 
dissociate Cooper pairs in the thin surface layer below T$_{\rm C}$, 
so that the PA wave travels from normal into superconductive phases. 
The incident light was loosely focused 
to the spot size of 1-mm diameter on the sample surface, and the intensity 
was kept at about 200 $\mu$J/pulse in order to avoid irradiation 
damage on the surface. 

The PA wave detected by the PZT was directly measured 
by an oscilloscope without any preamplifier. 
To examine the change of PA signal around 
T$_{\rm C}$, the PA signal was stimulated with chopped 
continuous-wave (cw) Ar-laser light of 2.41 eV and was picked up 
by a two-phase lockin detector. The incident light was loosely 
focused to the size of 2-mm diameter on the specimen surface, 
and the power was 10 mW. 

\begin{figure}
\includegraphics[width=8.0cm,clip]{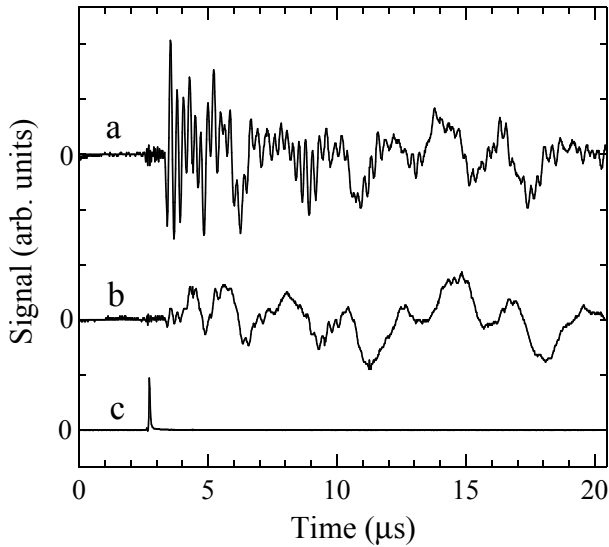}%
\caption{Curves a and b: PA waves at 4.2 and 28.8 K, respectively. 
Curve c represents temporal profile of incident laser pulse detected 
by a photodiode; the peak indicates the time 
at which the laser pulse reaches onto the sample surface. 
}
\label{fig2}
\end{figure}

\section*{RESULTS}
\begin{figure}
\includegraphics[width=8.0cm,clip]{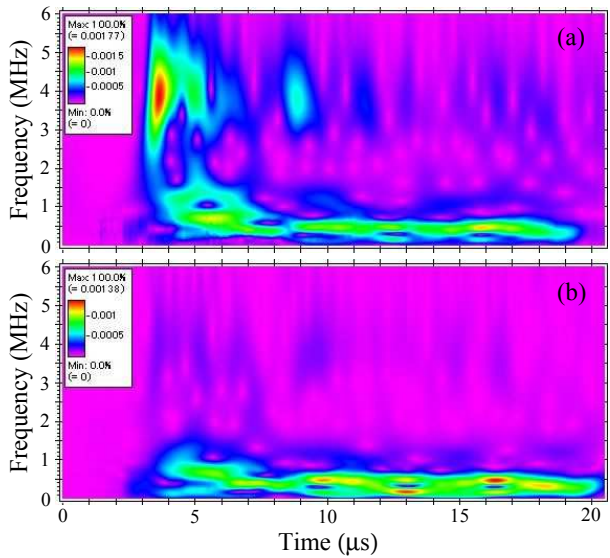}%
\caption{Time-frequency-domain image plot of PA wave in Fig.\ \ref{fig2}: 
(a) 4.2 K and (b) 28.8 K. The image maps were obtained by 
wavelet analysis. 
}
\label{fig3}
\end{figure}

Figure \ref{fig2} shows the PA waves at 4.2 (curve a) and 28.8 K 
(curve b) in the Pb single crystal. Curve c in 
Fig.\ \ref{fig2} displays the temporal profile of 2.33-eV and 5-ns 
laser pulse, measured by a photodiode. The incident laser 
pulses reached onto the sample surface at 2.72 $\mu$s. As seen in Fig.\ 
\ref{fig2}, the PA wave at 4.2 K includes many sharp spikes while the wave 
at 28.8 K has far less spikes. No further fine structure of PA wave 
is not observed by enlarging the waves at 4.2 and 28.8 K in the time 
domain. The difference of the two waves suggests that the PA wave at 4.2 K 
has a large amount of MHz components and is indeed 
presented in Fig.\ \ref{fig3} as the image plot in the time-frequency 
domain. Concerning with the shape of PA waves, it is to be noted 
that the PA wave measured by PZT's, in principal, includes the 
multiple reflection in the crystal and the ringing 
in the PZT simultaneously. The effect is well discriminated below 
by analyzing PA wave in the time-frequency domain. 

Figure \ref{fig3} is the result of time-frequency-domain analysis 
using wavelet~\cite{AGU} and presents the image plot of PA wave in 
Fig.\ \ref{fig2}. Wavelet transformation enables to extract the 
frequency component from wave in the time domain. The method has 
multiresolution and is a superior extension of Fourier transformation. 
Figure \ref{fig3}(a) corresponds to the result at 4.2 K 
(curve a in Fig.\ \ref{fig2}) and Fig.\ \ref{fig3}(b) to that at 28.8 K 
(curve b). Prominent signal appears at 4 MHz only at 4.2 K 
and the MHz component is strongly suppressed at 28.8 K. 
The peak position corresponds to the resonance of PZT and 
indicates the strong PA signal at MHz range. In this setup, a part of 
the strong PA signal appears prominently by the PZT resonance. 
On the other hand, oscillations are observed 
at 0.4 MHz in both image maps. The component at 0.4 MHz agrees 
with ringing frequency in the PZT;~\cite{Iwanaga_e} 
the ringing effect was detected in the configuration that 
the sample is removed in Fig.\ \ref{fig1}. Therefore, the PZT 
ringing is ascribed to the heat by laser irradiation. 
Presumably, the component at 0.4 MHz is induced by the thermal 
wave arrived at the interface between the crystal and PZT. 

\begin{figure}
\includegraphics[width=8.0cm,clip]{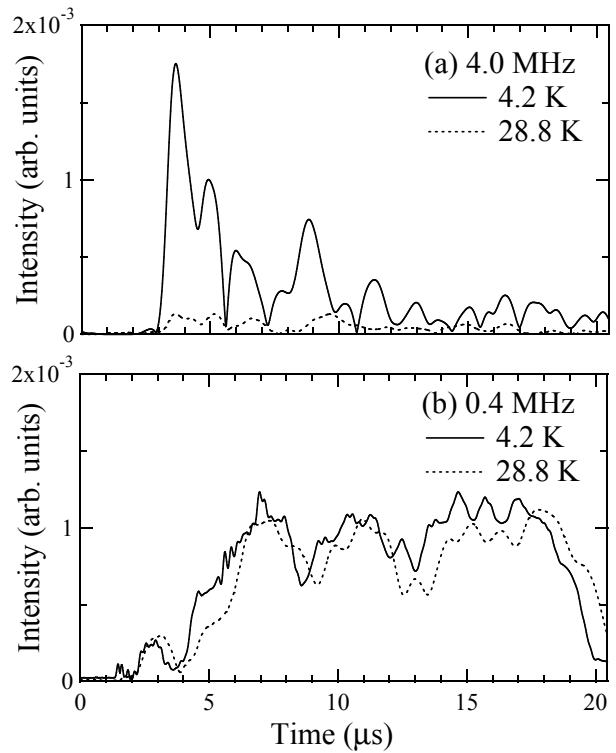}%
\caption{Temporal profile of time-frequency plot 
at (a) 4.0 and (b) 0.4 MHz. 
The profiles are extracted from Fig.\ \ref{fig3}. 
Solid lines represent the profiles at 4.2 K and dotted lines at 
28.8 K. 
}
\label{fig4}
\end{figure}

Figure \ref{fig4} displays temporal profiles at 4.0 and 0.4 MHz
in Fig.\ \ref{fig3}. It is apparent from Fig.\ \ref{fig4}(a) that 
PA signal at 4.2 K is enhanced at 4.0 MHz while 
Fig.\ \ref{fig4}(b) presents that the intensity and profile of PZT 
ringing are similar below and above T$_{\rm C}$. 
These results indicate explicitly that the two components are 
independent to each other; in other words, the ringing at 0.4 MHz 
is not induced by the MHz component. The likeness in 
Fig.\ \ref{fig4}(b) shows that the intensities of both waves are just 
proportional to the intensity of incident light and suggests that 
the ringing comes from thermal wave. 
Furthermore, the PA signal at 4.0 MHz grows rapidly at 3.0 
$\mu$s in Fig.\ \ref{fig4}(a) while the PA signal at 0.4 MHz 
increases gradually after 4.0 $\mu$s. 
The results also imply that the low-frequency component is 
induced by the wave different from the ultrasonic wave 
connected to 4.0 MHz component. Thus, 
taking account of the results in Figs.\ \ref{fig3} and \ref{fig4}, 
it is probable that the PA wave includes two different 
physical components, ultrasonic and thermal wave. 

\begin{figure}
\includegraphics[width=8.0cm,clip]{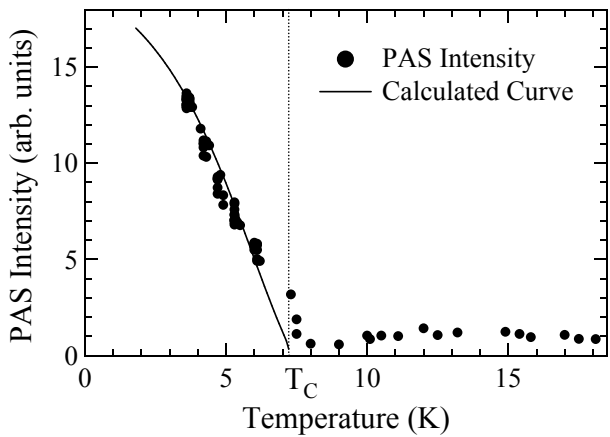}%
\caption{The intensity of PAS (solid circles) vs temperature. 
T$_{\rm C}$ denotes the superconductive transition temperature of 7.22 K. 
Solid line is calculated combining the attenuation of longitudinal 
ultrasonic wave with BCS theory, 
and fits the measured data below T$_{\rm C}$; 
the derivation of solid line is described in the text.
}
\label{fig5}
\end{figure}

In Fig.\ \ref{fig5}, the intensity of PA signal (PAS) is plotted with 
solid circle against temperature. 
The PA signals were stimulated with 2.41-eV, cw-laser light chopped at 
104 Hz and picked up from low to high temperatures. 
The intensity was measured with a two-phase lockin detector 
under the condition that each temperature is stable. 
The intensity keeps nearly constant from T$_{\rm C}$ to RT 
while it is enhanced below T$_{\rm C}$; the increase amount is almost 
in agreement with that of 4.0-MHz component in 
Fig.\ \ref{fig4}(a). Thus, the PA signal picked up by 
the lockin detector is ascribed to the leading component of PA wave. 
The temperature profile of PAS intensity is reproduced 
in Fig.\ \ref{fig5}; the calculated curve (solid line) 
is derived as follows. 

\section*{DISCUSSION}
Because of the leading component of PA wave and the present experimental 
configuration, it is assumed here that the PA signal in 
Fig.\ \ref{fig5} comes from longitudinal ultrasonic wave. 
Then, Fig.\ \ref{fig5} can be 
regarded as the plot of intensity of longitudinal ultrasonic wave. 
In this case, the ordinate is proportional to decayed intensity 
$\exp(-\alpha d)$, where $\alpha$ stands for 
absorption coefficient of the ultrasonic wave in the crystal and $d$ is 
thickness of the crystal ($d=1.0$ mm). In the superconductive phase, 
the $\alpha$ has to be replaced with $\alpha_s$ 
which is the absorption coefficient of longitudinal 
ultrasonic wave in the superconductive phase. 

To analyze the measured PAS intensity, the interface loss of 
signal has to be included. In fact, the PA signal decays in 
the crystal and moreover at the interface between the crystal and 
PZT. Therefore, an interface loss factor $a_{int}$ is introduced 
($a_{int} \ge 1$), and the measured PAS intensity is proportional to 
\begin{equation}
\exp[-a_{int}\alpha_s({\rm T}) d] 
\label{PASint}
\end{equation}
for ${\rm T} \le {\rm T}_{\rm C}$. 
The $a_{int}$ is treated as a fitting parameter below. 

The values $\alpha_s$(T) at 4.0 MHz are necessary in evaluating 
Eq.\ (\ref{PASint}). However, the values are not available in 
existent literature. Therefore, another procedure is chosen: 
(i) First, the $\alpha_s$(T) is evaluated by combining the absorption 
coefficient $\alpha_n$(T) in the normal state 
with the ratio $\alpha_s/\alpha_n$ derived by 
BCS theory.~\cite{Bardeen} 
Though the absorption coefficient $\alpha_n$(T) was reported only at 26.6 
MHz (Ref.~\onlinecite{Bommel}), the $\alpha_n$(T) can be evaluated 
around T$_{\rm C}$ from the literature because the frequency dependence 
is known and the temperature dependence 
is independent of ultrasonic frequency below tens of MHz 
(Ref.~\onlinecite{Morse}). 
Also, the ratio $\alpha_s/\alpha_n$ is 
written such as 
\begin{equation}
\alpha_s({\rm T})/\alpha_n({\rm T}) = 
2/\{1+\exp[\Delta ({\rm T})/k_B {\rm T}]\} 
\label{ratio:alpha}
\end{equation}
where $2\Delta ({\rm T})$ is the energy gap of superconductive state, 
which is expressed as $\Delta ({\rm T})=\Delta (0)\sqrt{1-({\rm T/T_C})^2}$ 
(Ref.\ \onlinecite{Bardeen}), 
and $2\Delta (0) = 4.38 k_B {\rm T_C}$ for Pb (Ref.\ \onlinecite{Kittel}). 
(ii) The PAS intensity is fitted by Eq.\ (\ref{PASint}) with changing 
the $a_{int}$. (iii) Finally, the most fitted value is 
searched by varying the proportionality constant of Eq.\ (\ref{PASint}). 

In the fitting procedure, after the $a_{int}$ is uniquely determined, 
the proportionality constant multiplied by Eq.\ (\ref{PASint}) 
is evaluated uniquely. Thus, the solid line in Fig.\ \ref{fig5} is 
obtained and seems to reproduce the data below T$_{\rm C}$ 
fairly well. 

The most fitted $a_{int}$ is estimated to be 4.1 by using the relation of 
$\alpha_n \propto \omega^2$. This is net interface loss and 
means that the PA wave is reduced at the interface. 
The resuction is suggestive of not optimized interface coupling. 

In the above analysis, the $\alpha_n$ is simply combined with 
the $\alpha_s/\alpha_n$. This assumes that the longitudinal ultrasonic 
wave is simply described by BCS theory and is not influenced by 
strong electron-phonon coupling in Pb. In fact, ultrasonic absorption 
coefficient at more than a-few-tens MHz deviates from 
the simple BCS result.~\cite{Deaton} However, as frequency becomes lower, 
the coefficients get close to the values derived from BCS 
theory.~\cite{Deaton} 
Therefore, the simple analysis using Eqs.\ (\ref{PASint}) and 
(\ref{ratio:alpha}) is found relevant to PAS of 4.0 MHz. 
Moreover, the analysis suggests that the normal state generated by 
photoexcitation gives little influence on PA wave, that is, 
the state is induced only in thin surface layer. 

As for the strong electron-phonon coupling, it is significant to 
measure transmission spectrum in the frequency domain by 
using calibrated PZT's. The PA wave is a kind of self-induced 
ultrasonic wave, and the transmission spectrum reveals 
the propagation mode; moreover, the mode includes 
the information on the electron--ultrasonic-wave interaction. 
The interaction has been classified with $q$ (wavenumber of ultrasonic wave) 
and $l$ (mean free length of electrons) phenomenologically. 
The transmission spectrum would enable to analyze $q \cdot l$ effect 
quantitatively in experiment. Indeed, the most effective $q \cdot l$ was 
estimated~\cite{Deaton} and corresponds to about 10 MHz; 
since the frequency is rather close to the present experiment, 
the transmission measurement seems realistic. The strong-coupling 
effect could be tested in detail by analyzing such transmission. 

As seen in this study, the PA wave in the normal state is regarded as 
thermal wave. The property of PA wave is perhaps common in normal metals 
because the attenuation of longitudinal ultrasonic wave is similar among 
them. In the case, the PAS detected by lockin equipment has to be 
analyzed on the basis of this property. 

Laser-induced acoustic waves have been reported.~\cite{Thomsen,Cheng} 
Since the comparison with the present PA study would attract an interest, 
a few comments are made here. 
The laser-induced acoustic waves are induced by ps- or fs-pulsed laser 
light and are extracted from the transient reflection detected 
with the pump-probe technique. Therefore, the acoustic signals have the 
frequency at GHz to THz and often correspond to optical phonons. 
The signals are induced at the laser-injected surface. 
The method is suitable to observe surface-layer phenonema because 
the attenuation of wave is typically 1 $\mu$m and the wave cannot travel 
through bulk samples. On the other hand, the present PA wave has the 
frequency at MHz 
and transmits over 1 mm. The MHz wave does not destroy Cooper pairs 
because the frequency is far smaller then the gap frequency 
determined by 2$\Delta$(T). That is, the PA wave travels in the 
superconductive phase. The frequency distribution of PA wave results from 
the ultrasonic propagation mode connected to energy dissipation. Therefore, 
the PA wave would provide new insights about the propagation mode and 
energy transport in the superconductive phase. 

In conclusion, PA waves have been explored in the time-frequency domain, 
so that it is clarified 
that the 4.0-MHz component highly transmits and the thermal wave 
is also observed as the PZT ringing in the superconductive phase, 
while the thermal wave is dominant in the normal state. The enhancement 
of PAS intensity is reproduced fairly well from the analysis based on the 
attenuation constant of longitudinal ultrasonic wave which satisfies 
the relation in BCS theory. Consequently, it is found that 
the PA signal below T$_{\rm C}$ is mainly composed of longitudinal 
ultrasonic wave in the present configuration. The enhanced frequency of 
PA wave probably comes from the propagation mode of ultrasonic wave 
in the superconductive phase though further measurement using caribrated 
PZT's is necessary to detemine the frequency distribution. 
The mode is likely associated with the 
effective interaction with superconductive electrons. 
The analysis of temperature-dependent PA-signal intensity suggests that 
the normal state hardly contributes to the PA signal, that is, 
the breaking of superconductive phase due to photoexcitation 
is restricted only to thin surface layer in the crystal. 
As a result, the PA wave propagates through the superconductive phase. 

\begin{acknowledgments}
I would like to appreciate the support for the PA measurement by 
Prof.\ T.\ Hayashi (Kyoto University). 
This study was supported in part by Grant-in-Aid for Research Fellow 
of the Japan Society for the Promotion of Science. 
\end{acknowledgments}

\bibliography{iwanaga12}

\end{document}